# Crossing the principle–practice gap in AI ethics with ethical problem-solving

Nicholas Kluge Corrêa[1,4] · James William Santos[4,5] · Camila Galvão[2,4] · Marcelo Pasetti[2,4] · Dieine Schiavon[2,4] · Faizah Naqvi[3] · Robayet Hossain[3] · Nythamar De Oliveira[2,4]

**Abstract**
The past years have presented a surge in (AI) development, fueled by breakthroughs in deep learning, increased computational power, and substantial investments in the field. Given the generative capabilities of more recent AI systems, the era of large-scale AI models has transformed various domains that intersect our daily lives. However, this progress raises concerns about the balance between technological advancement, ethical considerations, safety measures, and financial interests. Moreover, using such systems in sensitive areas amplifies our general ethical awareness, prompting a re-emergence of debates on governance, regulation, and human values. However, amidst this landscape, how to bridge the principle–practice gap separating ethical discourse from the technical side of AI development remains an open problem. In response to this challenge, the present work proposes a framework to help shorten this gap: *ethical problem-solving* (EPS). EPS is a methodology promoting responsible, human-centric, and value-oriented AI development. The framework's core resides in translating principles into practical implementations using impact assessment surveys and a differential recommendation methodology. We utilize EPS as a blueprint to propose the implementation of an *Ethics as a Service Platform*, currently available as a simple demonstration. We released all framework components openly and with a permissive license, hoping the community would adopt and extend our efforts into other contexts. Available in the following URL https://nkluge-correa.github.io/ethical-problem-solving/.

**Keywords** Artificial intelligence · Ethics · Principle–practice gap · Ethics as a service

## 1 Introduction

The late 2010s, especially after the beginning of the deep learning revolution [1, 2], marked a rising interest in AI research, underlined by an exponential increase in the academic work related to the field [3–5]. The confluence of technical advancements (i.e., breakthroughs in deep learning and increased computational power), data availability (i.e., the proliferation of data through social media, smartphones, and IoT devices), and massive investments (i.e., governments and private companies began investing heavily in AI R &D) played a crucial part in the expansion of the field [6–8].

✉ Nicholas Kluge Corrêa
kluge@uni-bonn.de

James William Santos
james.santos@edu.pucrs.br

Camila Galvão
camila.galvao@edu.pucrs.br

Marcelo Pasetti
paulo.pasetti@edu.pucrs.br

Dieine Schiavon
dieineb@ufcspa.edu.br

Faizah Naqvi
faizah_naqvi@brown.edu

Robayet Hossain
robayet_hossain@brown.edu

Nythamar De Oliveira
nythamar.oliveira@pucrs.br

[1] University of Bonn, Bonn, NRW, Germany
[2] PUCRS, Porto Alegre, Brazil
[3] Brown University, Providence, USA
[4] RAIES (Rede de Inteligência Artificial Ética e Segura) network, Porto Alegre, RS, Brazil
[5] Universität Erfurt, Erfurt, Germany

In the past few years (2022–2023), AI has entered an era where organizations release large-scale foundation models every few months [4], with systems like GPT-4 [9], Llama 2 [10], Gemini [11], Whisper [12], CLIP [13], among many others, becoming the basis of many modern AI applications. The capabilities of such models, ranging from human-like text generation and analysis to image synthesis and unprecedented speech recognition, have revolutionized the public consciousness of AI and transformed how we interact with technology.[1]

This accelerated progress comes with its own set of challenges. Notably, academia released the majority of state-of-the-art machine learning models until 2014. Since then, the market-driven industry has taken over [3, 4]. Big tech companies are the current major players in the research and development of AI applications. This shift leads us to question the balance between ethical considerations, safety measures, technological progress, and revenue shares. In other words, prioritization of revenue and "progress in the name of progress" may undercut ethical and safety concerns. Meanwhile, the use of AI systems in sensitive areas, such as healthcare [15, 16], banking services [17], public safety [18, 19], among others [20–22] prompted a re-emergence of the debate surrounding the ethical issues related to the use, development, and governance of these systems and technologies in general [23–26].

The AI safety field of research emerges as one possible solution to this unprecedented expansion. AI safety focuses on approaches for developing AI beneficial to humanity and alerts us to the unforeseen consequences of AI systems untied to human values [27–30]. At the same time, the growth of the research field in terms of regulation and governance demonstrates an apparent broad consensus on the human values (principles) relevant to AI development and the necessity of enforceable rules and guidelines to apply these values [3, 31–34]. Under this scenario, events like the "Pause Giant AI Experiments open letter" [35], among others [36, 37], are a symptom that even the industry recognizes that the unrestrained AI development impacts will not be positive or manageable [30]. With that realization, the outcry for regulatory input in the AI development industry grew stronger [38–40]. Ultimately, the debate on whether consensus on ethical principles exists, how to translate principles to practices, and the interests underlying the push for creating

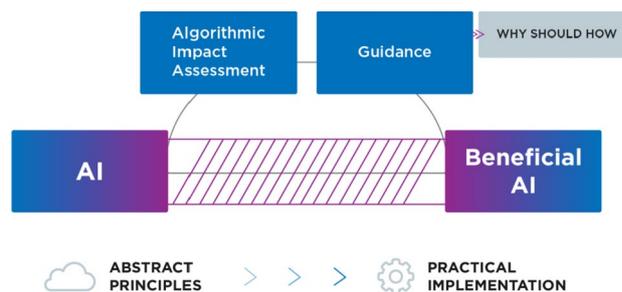

**Fig. 1** EPS seeks to bridge the principle-practice gap between principles and practical implementations, giving developers tools to use in the development cycle of an AI system. The general workflow of this method consists of an evaluation (Impact Assessment) and a Recommendation stage, structured in a WHY, SHOULD, and HOW format

normative guidelines presents critical unanswered questions around AI research [41].

This work seeks to tackle a critical point within these aforementioned unanswered questions. The principle–practice gap (i.e., translating principles to practice [42]) presents the sociotechnical challenge of managing the expectations of those who seek a magical normative formula and working with developers to foster design ingrained with ethical principles. With this in mind, we present ethical problem-solving (EPS), a method to aid developers and other professionals responsible for creating autonomous systems (Fig. 1).

Our framework's core resides in translating abstract principles into practical implementations through impact assessment and a differential recommendation methodology. The EPS builds and improves upon similar works, where proposals are usually limited to paper or worksheet frameworks, while also presenting novel contributions: an implementation of EPS tailored to Brazil's context. In the following sections, we will explore the challenges faced in ethical AI development, the path toward the EPS, its process, and resources. We released all components of our current implementation openly and with a permissive license, hoping the community can adopt and extend our efforts into other contexts.[2]

## 2 Related works

This section provides an overview of practical frameworks for AI ethics, specifically emphasizing approaches and methods that help translate ethical principles into developmental insight.

---

[1] In this study, the term Artificial Intelligence (AI), and more specifically, AI systems and applications, are defined as the products generated by the interdisciplinary field of computer science, cognitive sciences, and engineering that focus on creating machines capable of performing tasks that typically require human intelligence (e.g., natural language processing, problem-solving, pattern recognition, decision-making, forecasting, etc.) [14].

[2] All materials tied to this study are available in https://nkluge-correa.github.io/ethical-problem-solving/.

According to Taddeo and Floridi [42], creating methods to anticipate ethical risks and opportunities and prevent unwanted consequences is the crucial aspect of what the authors call translational ethics. One significant issue for translational ethics is the definition of the appropriate set of fundamental ethical principles to guide the creation, regulation, and application of AI to benefit and respect individuals and societies. Another is the formulation of foresight methodologies to indicate ethical risks and opportunities and prevent unwanted consequences (e.g., impact assessments, stakeholder engagements, and governance frameworks). In the context of AI ethics, translational ethics aims towards concrete and applicable practices that realize broadly accepted ethical principles. In other words, the translation of theoretical conclusions into adequate practice [43]. However, the authors do not present a particular foresight methodology in the translational framework proposed by Taddeo and Floridi. Nevertheless, they have an optimistic view towards a global convergence of ethical principles that could bring regulatory advancements, as shown in other fields of applied ethics.

Another approach that aims to assess the ethical aspects of AI systems is the VCIO-based (Values, Criteria, Indicators, and Observables) description of systems for AI trustworthiness characterization [44]. First introduced by the AI Ethics Impact Group in the "From Principles to Practice—An Interdisciplinary Framework to Operationalize AI Ethics" [45], the VCIO approach identifies core values, defines specific criteria, establishes measurable indicators, and utilizes observable evidence to measure the alignment of AI systems according to the selected values (i.e., transparency, accountability, privacy, justice, reliability, and environmental sustainability), which steam from a meta-analysis of relevant publications. The VCIO approach culminates in an AI Ethics Label. The label offers a rating for each value assessment proposed by the VCIO approach and later communicates an AI system's ethical score. However, no prerequisites exist to build a label based on the VCIO score. Companies, users, or government bodies can set the requirements for a minimum level of risk within this framework. Moreover, the following publication tied to the VCIO approach [44] clearly states that the VCIO's description/evaluation is independent of the risk posed by the technology under consideration and does not define any minimum requirements. It merely describes compliance with the specified values.

Even though VCIO is a valid contribution to the field, its limitations are also apparent in its description. First, the source of the values stated in the papers and the parameters of the meta-analysis that originated them are unclear. Despite being well-known principles in the AI ethics literature [31–34], we argue that not knowing how they are defined or uncovered represents a weak spot in the underlying methodology of the VCIO framework. Second, the self-imposing character of the risk evaluation requirements can be an issue and potentially undermine the effect of the approach. If there is no objective measure of the potential risks and impacts on ethical principles, the whole project of applying principles to practice may become moot.

Another approach that instrumentalizes ethical principles into palpable development tools is the Google People AI Research Guidebook [46], developed by PAIR's (People + AI Research) multidisciplinary team. The guidebook aims to empower researchers and developers to create AI systems that are fair, inclusive, and unbiased. The guidebook provides six chapters and their respective worksheets that acknowledge the potential risks associated with AI development and emphasize the importance of addressing bias, fairness, interpretability, privacy, and security. The guidebook encourages researchers to adopt a multidisciplinary approach incorporating insights from diverse fields. The worksheets provide strategies for understanding and mitigating bias, creating interpretable AI models, implementing privacy-preserving techniques, and promoting responsible data-handling practices.

PAIR's guidebook represented a commendable step towards building AI systems that are ethically sound, unbiased, and beneficial to all. Nevertheless, the guidebook and respective worksheets do not present a standard to evaluate whether there is progress in the development. Users of the methodology are left free to gauge their successes and failures, making the whole approach tied to the expectations of the user, who is also the evaluator. Also, the worksheets themselves do not present an approachable user interface.

Other examples of methods created to deal with the same issues previously mentioned works sought to tackle include:

- The Digital Catapult AI Ethics Framework [47] is a set of seven principles and corresponding questions to ensure AI technologies' responsible and ethical development. The framework advises the developer to consider the principles and their questions. The Digital Catapult underlines that the objective of the questions is to highlight the various scenarios in which ethical principles may apply to the project. Nevertheless, just like in the VCIO methodology [44, 45], their method needs to clarify the constitution of the set of relevant values. Meanwhile, the voluntary aspects of the approach rely on the awareness developers must have regarding AI's potential risks and ethical blank spots.
- Microsoft's AI Ethical Framework [48] is a set of principles and guidelines designed to ensure AI's responsible and ethical use. The framework encompasses six principles that should guide AI development and use: fairness, privacy and security, transparency, trust and security, inclusion, and accountability. Each (group of) principle(s) sets its goals and practical measures, bring-ing back the idea of pragmatizing ethical principles. By adhering to these principles, Microsoft aims to contribute to positive societal impact while mitigating potential risks associated with AI technologies. While the source of the ethical principles remains undisclosed in the paper (a recurring theme in the literature related to this work), the tools meant to achieve said principles are underrepresented, leaving much of the heavy lifting to the developers themselves.

- Morley et al. [49] investigation also revolves around the gap between AI ethics principles, practical implementations, and evaluation of the existing translational tools and methods. The authors propose an assistance method for AI development akin to a Platform as a Service (PaaS). PaaS represents a set-up where core infrastructure, such as operating systems, computational hardware, and storage, is provided to enable developers to create custom software or applications expeditiously. Meanwhile, Ethics as a Service (EaaS) seeks to contextually build an ethical approach with shared responsibilities, where the EaaS provides the infrastructure needed for moral development. In their work, Morley et al. mentions several components that should be a part of this kind of platform, for example, an independent multi-disciplinary ethics board, a collaboratively developed ethical code, and the AI practitioners themselves, among others. EaaS as an idea has the advantage of being relatable to modern tech companies, where organizations subsidize much of their work to specialized third parties. Also, having an ethical evaluation performed by a third party has merits. However, the actual operation of the EaaS framework and the content of the service provided are still ongoing research for the authors.

- Baker-Brunnbauer [50] proposes the Trustworthy Artificial Intelligence Implementation (TAII) Framework, a management guideline for implementing trusted AI systems in enterprises. The framework contains several steps, starting with an overview of the system to address the company's values and business model. Then, the focus shifts to the definition and documentation of the stakeholders and the exact regulations and standards in play. The assessment of risks and compliance with the common good follows. Ultimately, the framework should generate ethical principles suitable for translation into practical implementations, while executing and certifying the results should be the last step. However, two vulnerabilities of the TAII approach are its self-imposing nature and the many steps involved in the framework. Given these flaws, the iteration of the many measures proposed could leave evaluators blind to the weaknesses of their creation.

- As our last mention, we cite the framework created by Ciobanu and Mesnita [51] for implementing ethical AI in the industry. The proposed framework comprises AI Embedded Ethics by Design (AI EED) and AI Desired State Configuration (AI DSC). The AI EED stage is where the developer can train its model to address the specific ethical challenges of a particular AI application. Meanwhile, the developer or the consumer can define the relevant AI principles using the VCIO approach as the normative source. The AI DSC stage focuses on actively managing the AI system post-implementation through constant user feedback. However, it is unclear how the framework operationalizes its stages, except for the VCIO approach, which also has shortcomings. Also, the framework relies heavily on the interest and acute awareness of the general public to provide feedback on an end-to-end process where the AI system is adapted to the contextual reality where it is implemented.

As shown above, we can see many efforts to incorporate ethics in AI development and deployment. Many of these attempts are still extra-empirical and subject to improvement, modification, and actual deployment. Also, these attempts further justify our introduction's main point: the imperative necessity to anticipate and mitigate ethical risks in AI system development.

To build upon the work mentioned above, in the following sections, we propose a framework (EPS) that, we argue, is ethically and theoretically grounded, simple, practical, and not self-imposing.

## 3 Methodology

From the gaps in previous proposals and grounded in our critical analyses of the field of AI ethics, the EPS presents:

- A set of assessment surveys aimed at helping evaluators estimate the possible impacts of a given AI application.
- An evaluation matrix to classify an AI application's impact level according to AI principles grounded in empirical research.
- A recommendation approach customized to each impact level. These recommendations provide practical suggestions for enhancing the ethical standards of the evaluated system or application.

Before diving into the implementational aspects of our method, let us first revise the philosophical roots of ethical problem-solving.

### 3.1 Theoretical grounding

The EPS is grounded on Critical Theory's social diagnosis methods and emancipatory goals, particularly Rahel

Jaeggi's "Critique of Forms of Life". [52]. That said, this work's approach to normativity is not focused on the more traditional aspects of normative theory (i.e., how to judge a particular action as moral and which parameters to use) but is aligned with the notion that normativity is created and reproduced through social practices as an ongoing process. Hence, Jaeggi's approach fits with normativity as an integral part of sociality, which cannot be dissociated from it to judge what is right or good. This argument is in line with her approach to immanent critique, and it resonates with a few authors who methodologically sustain immanent criticism as the path to avoid universalistic or relativistic tendencies [53–56]. However, what differentiates Jaeggi's approach from others is her conceptualization of forms of life, her normative recuperation of Hegelian theory, her problem-solving approach, and the open practical questions it leaves us with.

Jaeggi draws on the concept of "forms of Life" introduced by Ludwig Wittgenstein [57], which broadly refers to how individuals and communities organize their activities and ways of understanding the world. Jaeggi reformulates this concept to analyze the social and cultural structures that shape human existence, examining how they might limit human flourishing, autonomy, and self-realization. Jaeggi understands forms of life as clusters of social practices addressing historically contextualized and normatively defined problems. Ultimately, what shapes human existence is how exactly these problems are solved within a form of life.

Broadly speaking, the practices that constitute a form of life are connected in practical-functional ways. Some have a tangible sense of how functionally interdependent they are, such as agricultural practices required for urban consumption. In contrast, others do not, such as playing with children as an essential part of parenthood. In this sense, practices bring their interpretation as something (descriptive) and the functional assignment as being good for something (evaluative) in correlation with each other. Jaeggi thus understands forms of life as being geared to solve problems because their description already carries a meshed functional and ethical perspective. In other words, forms of life always entail an inherent evaluation, excluding pure functionality in human activities. The critical theory of technology, also known as Science and Technology Studies (STS), sustains a similar position that technology is value-laden like other social realities that frame our everyday existence [58, 59]. However, the argument of technology being permeated by values or biases in its inception only partially resonates with the proposition of this work. It is Jaeggi's proposal of the value-laden argument setting up a normative foundation and a problem-solving approach that fits the present endeavor.

Jaeggi asserts that the normative dimensions of forms of life are not static but rather rooted in a triangular relationship involving the current empirical state (is), normative claims (ought), and changing objective conditions. The normative claims reflect the expectation of particular manifestations of social practices as they developed historically, and the current empirical state reflects the actual state of social practices against the expectations and facing objective conditions. This continuous process of dynamic normativity can be illustrated by the development of the rural-feudal extended family into the bourgeois nuclear family due to changed socioeconomic (objective) conditions demanding changes in normative expectations. Further developments, ranging from patchwork families to polyamorous relationships, could also be understood as reactions to the new objective conditions now posed in turn, but not solved, by the bourgeois family [52].

To address the unsolved issues of social practices and ever-changing objective conditions, Jaeggi proposes problem-solving in the form of a hermeneutic anticipation (i.e., a recommendation) of an assumed solution (i.e., of a desirable goal) and the validity of such a recommendation can only be determined after addressing the identified problems. As we can see, Jaeggi's approach foreshadows a dynamic normativity that renews itself through iteration without necessarily implying progress from the outset, considering that the success or failure of the recommendation determines the evaluation of the addressed problem.[3] Jaeggi's proposal also raised criticisms regarding her conceptualization of forms of life, which was deemed exceedingly vague [70]. Also, despite Jaeggi's push to include some notion of social reality in her theoretical proposition, it seems that Jaeggi needed to go further as the approach lacks a clear connection with sociality [71, 72]. Although Jaeggi's work does open practical avenues, much of the work toward changing practices is left out of her proposal.

At this point comes the inspiration for the EPS; at first glance, it might seem striking that there is a slight difference in scope between Jaeggi's theoretically proposed criticism for societies and the EPS, which seeks to bridge values toward actionable practices of ethical AI development. It is worth underlining that the EPS is not attempting an overarching criticism of technology. However, it can still take advantage of a problem-solving approach that tackles a complex conceptualization and responsively grasps normativity. The EPS puts into practice the problem-solving process theoretically proposed on a different scale but with a complex subject matter and ethical stakes nonetheless. There is no doubt that AI systems involve ensembles of directives to solve diverse issues and that much of the inner workings can be elusive to our current understanding of the subject, not unlike the elusive character of our social practices and the historical background that supports them.

---

[3] Jaeggi's contribution to Critical Theory with the reclamation of Hegelian normative touchstones, drawing on Dewey [60, 61], MacIntyre [62–64], and Pinkard [65], sets itself apart from the deconstructive [66, 67] and the overarching constructivist [68, 69] branches of the tradition.

Nevertheless, despite the elusive character of the subject, there is an apparent demand for normalization or, at least, to understand how suitable norms should arise. To this point, the EPS utilizes the dynamic normative approach to test Jaeggi's theoretical proposal further. The EPS gauges the empirical state of systems (through survey assessment) to trace if their normative assignments are aligned with the ever-changing landscape of AI ethics (considering the state-of-the-art in the field). If there are discrepancies, then problem-solving takes over to align the system. The EPS shines on problem-solving because it enacts something only theorized by Jaeggi; it enables a clear connection between dysfunctionality and its normative assignment to reframe the current state of an AI system with the normative expectations of the field of AI ethics through its practical recommendations. Ultimately, the EPS adaptation of Jaeggi's dynamic normativity and problem-solving approach transforms the principle–practice gap into an ongoing task open to correction and responsive to its surrounding ethical field.

### 3.2 Finding values: a descriptive analysis of the field

Much like the previous works mentioned [45–48], we embarked on the essential task of surveying the landscape of AI ethics to identify its relevant values. However, one factor that distinguishes our framework from the prevailing body of literature rests in the work of descriptive ethics that preceded the development of EPS, where we rooted the principiological foundations of our framework through a descriptive analysis of how the field defines, instrumentalizes, and proposes AI principles worldwide. This descriptive work is entitled Worldwide AI Ethics (WAIE) [3].

For starters, WAIE draws inspiration from earlier meta-analytical works [31–33, 73, 74] and meticulously surveys a wide range of ethical guidelines related to AI development and governance through a massive effort of descriptive ethics implemented as a data science venture. In it, 200 documents are thoroughly analyzed, including private company governance policies, academic statements, governmental and non-governmental recommendations, and other ethical guidelines published by various stakeholders in over 30 countries spanning six continents. As a result, WAIE identified 17 resonating principles prevalent in the policies and guidelines of its dataset, which now provide the principles we utilize as the basis for the succeeding stages of the EPS, from the assessments to the recommendations.

Besides the fact that EPS stems from our own meta-analysis of the field, we argue that our principiological foundation differs in the following ways from other works:

1. WAIE uses a worldwide sample of 200 documents, a more diverse representation of global ethical discourse around AI than previous studies.
2. WAIE delivers its information in a data visualization way that is interactive and searchable and allows the study of correlations.
3. WAIE is granular, presenting a series of typologies that increase the insight users can gain.
4. WAIE is open source, allowing users to replicate and extend our results.

By employing the WAIE review to sustain the EPS methodology, we offer a more nuanced and empirically substantiated perspective on the ethical underpinnings of artificial intelligence, thereby enhancing the depth and rigor of our axiological basis.[4]

However, while the utilization of WAIE has undoubtedly provided a valuable foundation for descriptive, and now normative, ethics in AI, we must emphasize the significance of augmenting our value analyses with a critical evaluation of the risks associated with recently released AI models [9, 10, 10, 12, 13, 75, 76]. As already stated by Bengio et al. [30], the field's dynamic and rapidly evolving landscape necessitates constant vigilance in assessing emerging technologies' potential pitfalls and challenges (e.g., disinformation, algorithmic discrimination, environmental impacts, social engineering, technological unemployment, intellectual fraud, etc.). Hence, we augmented the development of EPS with a risk analysis of large-scale models released in the last 5 years [77]. Again, we released all materials tied to this analysis as an open and reproducible project.[5]

By incorporating a critical evaluation of values and known risks, we not only provide a more holistic perspective on AI ethics but also equip stakeholders with a timely understanding of the complex ethical considerations surrounding the deployment of AI systems. This integrated approach ensures that our work remains forward-looking and responsive to the ever-changing landscape in the AI field. These are all fundamental aspects for implementing our method, which, in the first instance, is supported by extensive descriptive work. An empirical-descriptive grounding is paramount to any attempt to pragmatize ethics.

However, this descriptive work only serves as a starting point. As already pointed out by Whittlestone et al. [78] in their critique of the proliferation of AI principles, we must be ready to realize that, by themselves, principles and risks are insufficient to guide ethics in practice. Hence, now that we have made clear the philosophical, axiological, and descriptive roots of our work, we showcase how we translated these into a practical framework for AI development in the following sections.

---

[4] All materials tied to the WAIE review are available in https://nkluge-correa.github.io/worldwide_AI-ethics/.

[5] Available in https://github.com/Nkluge-correa/Model-Library.

### 3.3 Defining risk and impact with algorithmic impact assessment

The first step in the EPS methodology is to gauge the state of a particular system via an impact assessment survey. Our decision to utilize an impact assessment approach comes from the bourgeoning landscape of legislative efforts worldwide currently focused on this topic (i.e., European Union [79], Brazil [80], the United States of America [81, 82] [83], several African states [84–86], Australia [87], Argentina [88], Egypt [89], Japan [90], Israel [91], Estonia [92], Peru [93], China [94], Russia [95], United Kingdom [96], Canada [97], among many others). Whether still in production or already enacted, the realization that governments should legally regulate AI is trending toward unanimity. For example, Brazil (the context where EPS came to be) does not have a bill regulating AI systems specifically. However, several bills to govern such technologies are currently the subject of debate in the National Congress. Nevertheless, from our analysis of these aforementioned regulatory efforts, we argue that two main trends are evident:

1. Determining the fundamental ethical principles to be protected (which we have achieved through the WAIE review).
2. A risk-based governance approach toward autonomous systems.

For disambiguation purposes, a risk-based governance approach implies a differential treatment concerning AI systems, i.e., different types of systems demand differential treatment pertaining to the risks they pose. For example, a spam filter would not require the same level of auditing and ethical concern as an autonomous vehicle.

In the EPS framework, we argue that the concepts of risk and impact concerning AI ethics differ in the *ex-ante* and *ex-post* relationships. Risk refers to the likelihood of negative consequences arising from the deployment and use of artificial intelligence systems, assessing the potential harm that could result from a particular AI application (ex-ante). On the other hand, impact refers to the magnitude and significance of the actual damage or benefit that occurs when these risks materialize or are mitigated, considering the real-world effects of AI systems on individuals, society, and the environment (ex-post). Regardless of their differences, both concepts are related to the impact these technologies can have on the wild.

Even though both of these terms are used interchangeably in many situations, we choose to use the term impact, aiming to encompass the potential problems (risks) and the actual consequences of harmful AI technologies, including their ethical, social, legal, and economic implications. At the same time, the term "impact" assessment is already accepted and used by the literature [98–100]. We argue that choosing the term "risk" assessment could entail only a preemptive approach toward assessing AI systems' negative impacts while introducing less-used terminology. Finally, we also point out that many of the known negative impacts of AI systems are currently documented in the form of "impacts" in many publications that present ethical assessments [101–106].

Another important aspect related to the development of impact assessment methods is that the impact of AI technologies can vary significantly depending on the cultural, social, political, and economic contexts. For example, concerns for the indigenous population must be considered sensitive topics in contexts such as countries that were former subjects of colonial rule. These topics are not quite paramount in the global north. An AI system that is thoughtfully designed concerning local culture, laws, socioeconomic conditions, and ethics is more likely to succeed and less likely to cause harm in our diverse global landscape [107, 108]. Hence, accounting for context is a critical step in the EPS framework to ensure a context-sensitive evaluation. As our theoretical grounding session stated, context is paramount to tracing and addressing the normative assignments in our daily practices. In our results section, we will showcase how we executed this in our implementation.

Finally, in our conception, the evaluation stage of any framework entailed in auditing AI systems should not be fully automated. The sole purpose of an evaluation should be to aid and inform a team of human evaluators, e.g., an ethics board, that can use such information to make a normative decision (e.g., a medical ethics board). In other words, we do not agree that ethical evaluations concerning human issues should be a matter of non-human deliberation. The development of this kind of evaluation board is beyond the scope of this study. Our only input is that such a group should be formed in the most interdisciplinary way possible, having representatives of all areas concerning AI ethics, e.g., Computer Science, Philosophy, Law, etc, as already suggested by other works [44–46, 109, 110]. Also, it is essential to note that in our conception, such evaluation should be administered by third parties or sources outside of the developmental cycle of the technologies under consideration, making the EPS framework not a self-evaluation method but a process that requires the collective engagement of AI developers and auditing parties.

### 3.4 WHY–SHOULD–HOW: a three-step approach to ethical problem-solving

The EPS framework acts as the bridge between the recognition of AI system dysfunctionality and its normative assignment. As previously mentioned, through EPS, it becomes possible to identify, assess, and understand the ethical implications of an AI system. Moreover, EPS provides practical recommendations to reframe an AI system's actual condition in alignment with the expectations of the AI ethics community. Therefore, problem-solving becomes the act of providing recommendations informed by the current state of the AI system and the normative assignments demanded from within the system and by the field of AI ethics. The WHY–SHOULD–HOW methodology appeared as a palpable and direct form to underline the relevancy of ethical principles in AI development while stating the normative standards and offering comprehensive recommendations to address the issues. This three-step process attempts to grasp the essential features of shortening the principle–practice gap in AI development.

First, the WHY component serves as the foundation, demonstrating the relevance of the AI principles to the specific issue at hand. It encourages practitioners and organizations to acknowledge why they should uphold particular values and the consequences of the opposite. This step is crucial as it sets the stage for a deeper understanding of AI applications' implications and broader societal and ethical impacts. Also, the WHY step is aligned with the epistemological claim that informed developers are better than uninformed ones, which revolves around the fundamental idea that ethical knowledge and understanding are essential to technological development. In this context, "informed" developers would better understand the principles, best practices, and technologies relevant to their field, while "uninformed" developers may lack this knowledge. In other words, explaining why something is suitable is the first step in any approach that seeks to promote moral reasoning convincingly. Otherwise, starting with an imperative claim may seem authoritative.

While the WHY step represents the foundation and contextualization of principles, SHOULD and HOW are its pragmatization. The last two stages are associated with different levels of impact, inferred at the EPS assessment stage (low, intermediate, and high).[6] The SHOULD aspect outlines the necessary steps to tackle ethical problems. By necessary, we mean that the measures indicated in this step represent the axiological content of each principle in the framework and are integral to developing an ethically attuned AI system. We can also define the SHOULD stage as an implementation of normative ethics in an applied format, where besides defining explicit "oughts," we stipulate criteria to help users and evaluators assess the compliance of a given system. This step traces a causal relationship between principles and observables, taking the VCIO approach as inspiration [44, 45].

However, the EPS aims to go beyond the normative guidance, presenting the how-to-do, i.e., the practical step. Hence, the HOW component offers valuable tools and strategies to implement the ethical recommendations in the SHOULD stage. In short, it equips developers, researchers, and organizations with the means to put ethical principles into practice. The scarcity of practical tools to address ethical matters within AI is a significant and concerning gap in AI ethics [3, 32, 111, 112]. Most of the literature that brings ethics to the development of applications does so through its descriptions of principles and extensive flowcharts of how the AI development process should be, failing to provide the practical support to address ethical problems or achieve the principles it underlines. In the meantime, developers often face unique and context-specific dilemmas, and without practical guidance, they may resort to ad-hoc solutions or bypass critical considerations altogether. Without available tools to guide developers or the know-how of how to use them, it is no surprise that there is a lack of standardized ethical practices in AI development [113, 114], resulting in blank spots and inconsistencies across the life cycle of AI projects.

Thus, given the deficits mentioned above, the normative step alone is insufficient to cross the principle–practice gap, making the absence of the HOW to step a clear blank spot in other works that our framework seeks to surpass. Ultimately, the WHY–SHOULD–HOW approach culminates in an educational step, acknowledging that responsible AI development is far from standard practice in the curriculum of many STEM-related fields that sprout most AI developers. In our envisioned form, this whole process aims to integrate professionals from humane sciences to STEM-field areas, and vice-versa, bringing developmental focus to ethics and developmental mindset to ethical considerations. Hence, if we suppose, taking a Virtue Ethics stance, that moral behavior can only stem from practice, our proposed framework allows practitioners to develop their virtues through training, which is why we implemented the HOW step as an "educational" step, besides a practical one.

In the following section, we delve into the heart of our work, presenting an implementation of the ethical problem-solving framework. This implementation offers a blueprint for constituting an EaaS platform to apply our envisioned framework.

---

[6] Increasing levels of impact demand additional recommendations and more severe implementations.

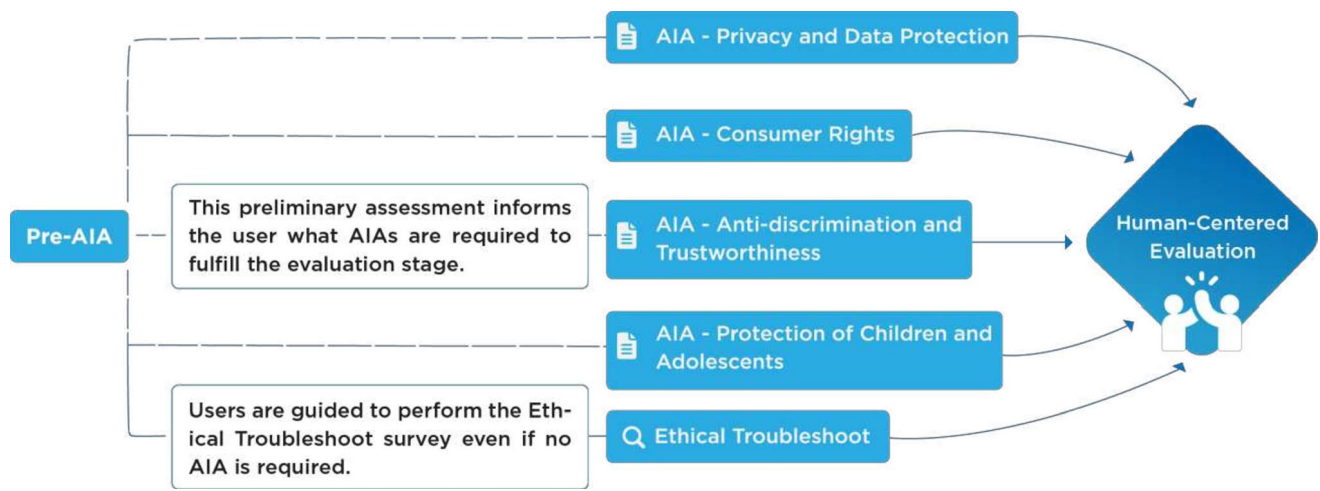

**Fig. 2** This flowchart illustrates the evaluation structure of the EPS framework. The dotted lines trace the pre-assessment leading through the algorithmic impact assessment surveys. The straight line represents the indispensable survey throughout the framework: the Ethical Troubleshoot. All survey assessments lead to a human-centered evaluation process

## 4 Results

In this section, we will show an implementation of the EPS framework. The idea of Ethics as a Service guided the creation of this implementation. Following the attempt of Morley et al. [49], this EaaS implementation would provide the infrastructure for ethical AI development akin to what a Platform as a Service offers, i.e., a platform where developers can submit their systems to ethical auditing.

As mentioned before, the EaaS idea has the advantage of being relatable to modern companies, where third parties constantly subsidize services and infrastructure that is too costly to maintain. While this may not be the case for large organizations (i.e., organizations that, besides having their own technological infrastructure, also have their own ethics boards), an EaaS may as well be a valid tool for companies that cannot sustain or afford their own AI ethics and safety auditing. Also, we again stress the value of a neutral, third-party auditing platform, regardless of the organization's size.

The components of our envisioned implementation are:

1. The EPS framework (questionnaires, evaluation metrics, recommendations, and educational aid).
2. A platform to apply this methodology.
3. An ethics board to perform the evaluations.

The subsections below present a step-by-step implementation of the EPS framework as an EaaS platform tailored to the Brazilian context.

### 4.1 Evaluation stage: algorithmic impact assessment and ethical troubleshoot

The flow of the ethical problem-solving framework begins with a pre-algorithmic impact assessment. The pre-assessment gauges preemptively the realm of impact of a particular system, leading to the actual tools of impact assessment. In other words, this preliminary assessment informs the user what algorithmic impact assessment surveys (AIAs) are required to fulfill the evaluation stage. For example, the user must perform the privacy and data protection AIA if the intended application utilizes personally identifiable information. Hence, after this brief assessment, the user is directed to the next step: the EPS' algorithmic impact assessment surveys (Fig. 2)

The algorithmic impact assessment surveys consist of questionnaires with pre-defined questions and answers that can be single-choice or multiple-choice (Fig. 3). Our current implementation of these covers the following themes: data protection and privacy, protection of children and adolescents, antidiscrimination, and consumer rights. Choosing these areas was a strategic move to gather resources since they are rich in legislative content in the Brazilian context. Even though Brazil still needs specific AI regulations, other legal sources can still be used to determine what is and is not allowed regarding the use and development of AI technologies. This design choice also highlights another important aspect of the EPS framework: the importance of contextual information while developing the evaluation stage of an implementation of our framework. This implementation aims to show stakeholders and

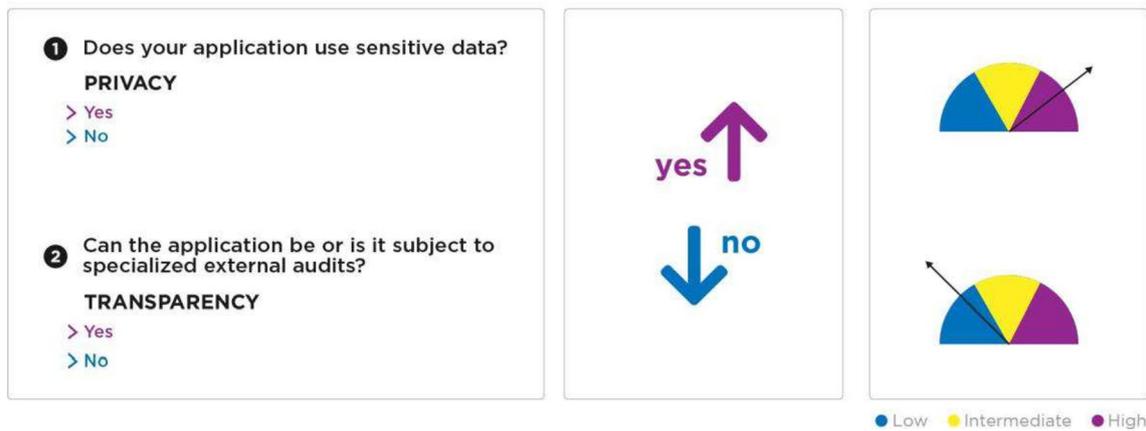

**Fig. 3** We developed the questions from the algorithmic impact assessment surveys to infer the level of impact a particular system may have on different ethical principles. Each question's response could either raise the impact, remain unchanged, or lower it if mitigating measures have been found

researchers how these assessments can be created even if AI is not explicitly regulated.

The questionnaires for the algorithmic impact assessment surveys entail that each of the questions identifies the AI's compliance with at least three ethical principles identified by the WAIE review. Hence, each of these generates impact scores relative to these assessed principles. As a result, distinct principles may serve as the basis for each question in each AIA. At the same time, each question's response could either raise the impact, remain unchanged, or lower it if mitigating measures have been found. In other words, we use objective questions tied to contextually relevant legally binding norms intended to guarantee a good life to infer the impact caused by a technology under consideration. Our current implementation of these impact assessment surveys presents for each question:

1. Their possible answers.
2. The scores related to each answer.
3. The principles impacted by the answers to each question.

Ultimately, these assessments can generate a standardized impact level on each ethical principle evaluated by each AIA. At the same time, the overall cumulative impact of all assessed principles represents the general impact of a system against a specific AIA. For example, in our privacy and data protection AIA, the following principles could be impacted, depending on the answers given by the user: privacy, transparency, and accountability. Hence, the final result of the privacy and data protection AIA presents an individual impact score for each principle and an overall score on the AIA itself (the standardized summation of each evaluated principle):

$$\text{Impact Score}_{privacy} = \frac{\text{Score}}{\text{Max Score}}$$

$$\text{AIA Score} = \frac{\text{Impact Score}_{privacy} + \text{Impact Score}_{transparency} + \text{Impact Score}_{accountability}}{\text{No of evaluated principles}}$$

The algorithmic impact assessment surveys use legally binding standards to deduce the implications of AI systems through an objective lens. However, these questionnaires provide an impact score that cannot address all of the ubiquities attached to the ethical issues that AI systems present. Hence, in our current implementation of the EPS framework, we developed a more qualitative survey to accompany them, entitled Ethical Troubleshoot, aimed at going beyond an objective evaluation.

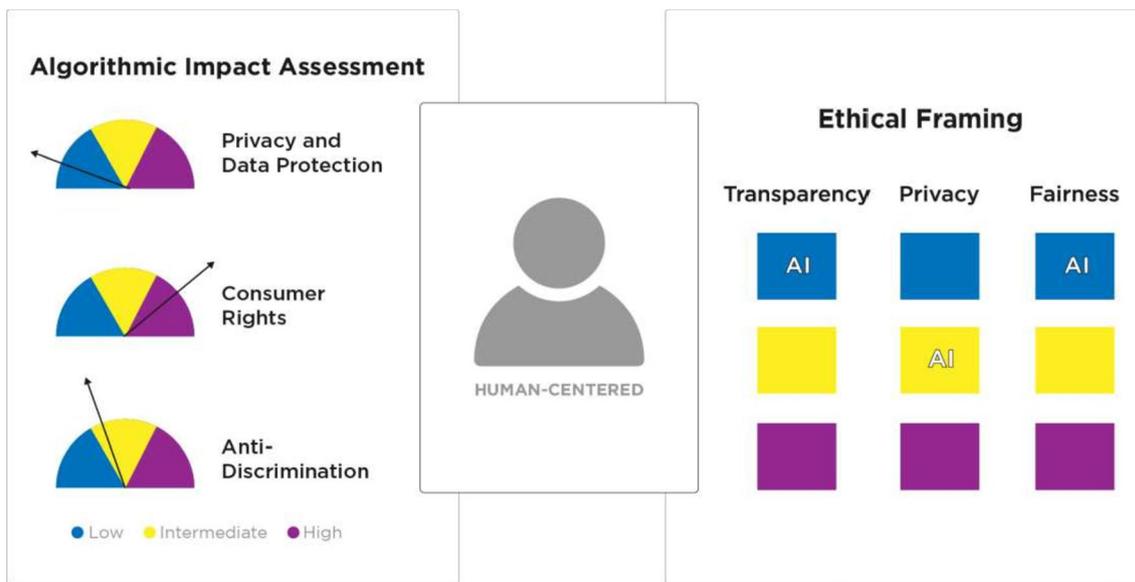

**Fig. 4** The EPS framework's evaluations are designed to help human evaluators assess the level of impact of an AI system. Each evaluated principle has three distinct levels of impact. After being informed by the outputs of the evaluation stage, human evaluators prescribe the particular impact level of an AI system regarding the ethical principles being considered

In short, the Ethical Troubleshoot survey seeks to allow the respondent to divulge how the development of a given AI system or application has been done in a human-centric way, e.g., how the needs of the intended users have been considered. It utilizes a combination of multiple-choice, single-choice, and open-ended questions to gauge the system's scope, its intended and unintended uses, and its target audience. We argue that a mixture of objective evaluation modes and more qualitative assessment forms can only augment a human-centric ethical evaluation (Fig. 4). In other words, this qualitative survey is meant to capture the aspects that the more objective and rigid AIAs could not assess. Our implementation of the Ethical Troubleshoot survey was mainly achieved by reverse engineering the VCIO method [44, 45] and the Google-PAIR worksheets [46].

As already mentioned, the sole purpose of these evaluation surveys is to help inform a team of human evaluators, e.g., an ethics board, that can use such information to make a human-centered evaluation.[7] The output of this decision is the recommendation stage.

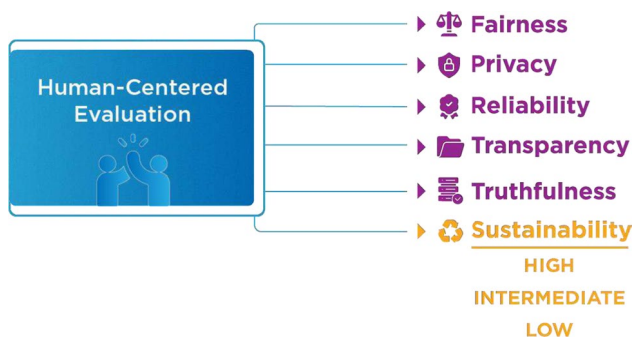

**Fig. 5** After being moderated and revised by an ethics board (human-centered evaluation), the assessment output is an ethical framing, where the system under consideration is classified with an impact level (high, intermediate, and low) for each of the evaluated principles

### 4.2 Recommendation stage: WHY–SHOULD–HOW

After the evaluation stage, the EPS framework requires that human evaluators classify the system under consideration in an impact matrix. The matrix comprises three recommendation levels tailored to each impact level—high, intermediate, and low—and six ethical principles gathered from the WAIE review, i.e., fairness, privacy, transparency, reliability, truthfulness, and sustainability. Hence, each principle has three distinct possible recommendations tailored to specific impact levels, e.g., Sustainability-low, Sustainability-intermediate, and Sustainability-high (Fig. 5).

---

[7] We define "human-centeredness," also known as human-centric design or user-centered design [115], as a methodology employed across various disciplines to prioritize human needs, behaviors, and preferences in creating and optimizing products, services, and systems. Hence, by a "human-centered evaluation," we here refer to a process of ethical deliberation and reasoning that requires human involvement and not automated processes.

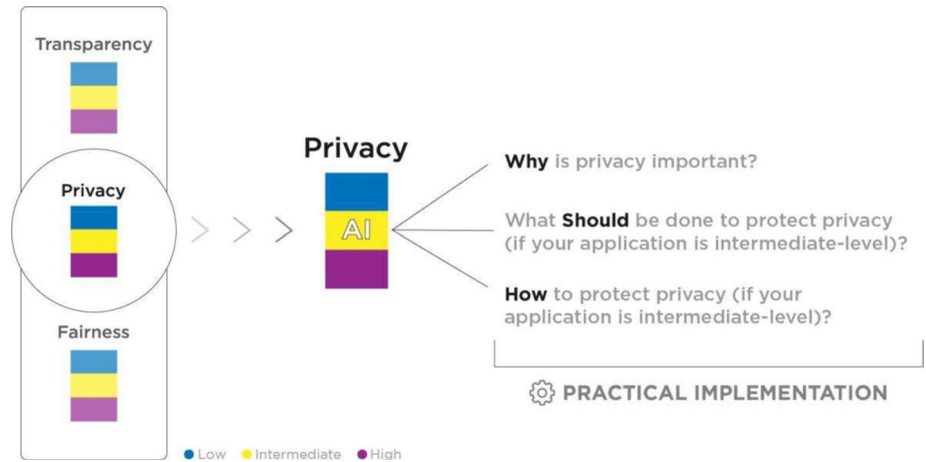

Fig. 6 Each level of recommendation regarding the principles utilized is structured around the WHY–SHOULD–HOW method. This allows the evaluators to make differential recommendations based on each principle's inferred level of impact. Subsequently, each level of impact presents differential recommendations with tailored tools and practices for that specific impact level

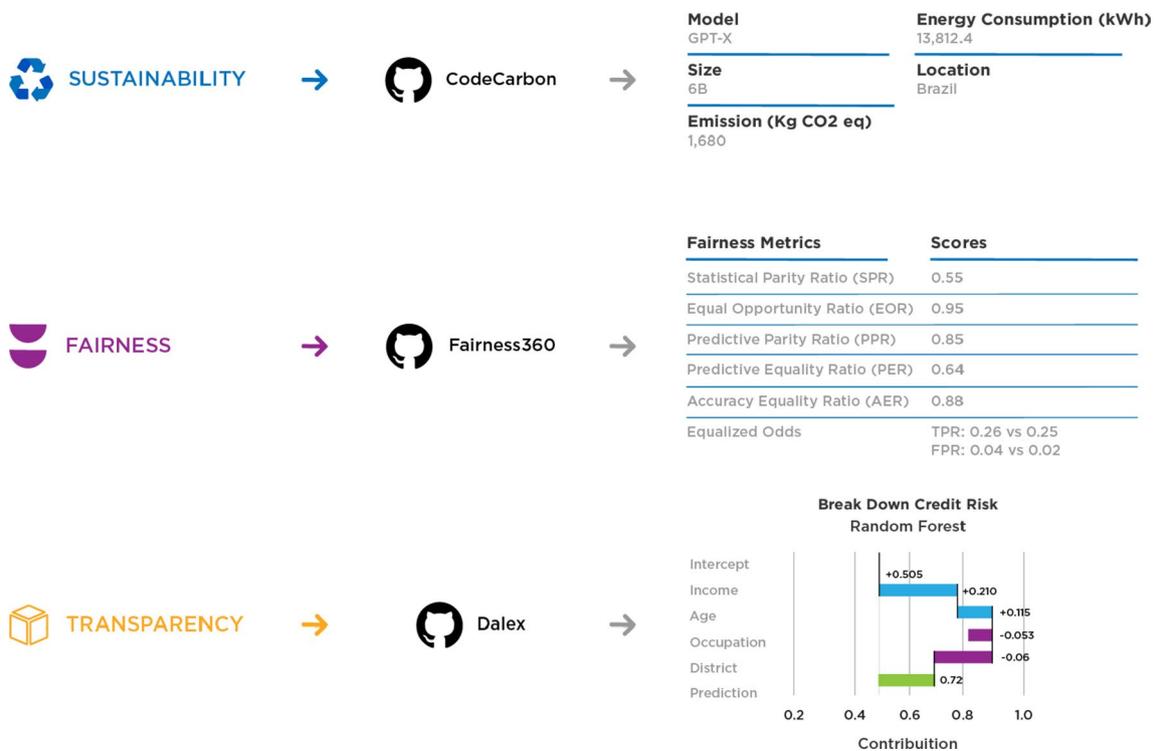

Fig. 7 We structured the HOW step around the question, "*How can a developer increase the robustness of a system in regards to a specific ethical principle?*", laying down step-by-step instructions, toolkits, educational resources, testing, and training procedures, among other resources that can help apply the principles used in the EPS

The WHY–SHOULD–HOW method is the format in which our approach presents the evaluation's outcome. First, the WHY step is structured to demonstrate the relevancy of each principle, providing the conceptualization and highlighting paradigmatic cases of deficit implementation in a structure that answers the questions "*What is said principle?*" and "*Why should you care about it?*". Second, the SHOULD step provides the metric utilized to gauge the level of recommendation regarding the corresponding principle, the level of recommendations indicated for the specific case, and the set of recommendations in a summarized form. Third and finally, the HOW component offers the practical tools and strategies required to implement the recommendations made in the SHOULD stage, i.e., it pragmatizes the normative recommendations of the previous step while also providing how-to instructions on using them (Fig. 6).

We developed the EPS framework to address the principle–practice gap explicitly. Given that we wished to go beyond simply "pointing to what tools developers can use," we developed an open-source repository containing a collection of demonstrations of how to use the developmental tools we recommend as part of the EPS [116]. This repository has many examples of tools and techniques developed to deal with the potential issues of an AI application (e.g., algorithmic discrimination, model opacity, brittleness, etc.), using as test cases some of the most common contemporary AI applications (e.g., computer vision, natural language processing, forecasting, etc.) (Fig. 7).

The effectiveness of a tool or framework often hinges on its ease of use and implementation. For instance, PyTorch's popularity in deep learning stems from, besides the inherent value of its automatic differentiation engine, its comprehensive documentation, which facilitates widespread adoption by lowering the entry barrier for newcomers to the field, simplifying complex procedures, like neural network engineering and training, to simple how-to-do examples. In the case of the EPS, we try to accomplish the same for the practices related to AI ethics and safety. To better exemplify this, let us describe a hypothetical use of our framework as an EaaS platform:

> **Hypothetical use case:** An organization in Brazil is in the process of developing an AI-empowered product. Before deploying it to its first users, the organization applies the EPS method via an EaaS platform to access ethical and legal compliance. During the evaluation stage, the organization answers the surveys in the best way possible, disclosing all information required, protected by a non-disclosure agreement between both parties. After the evaluation stage, the ethics board working behind the platform receives the results of the product evaluations. This information is also disclosed to the organization since it gives valuable information about the legal compliance of the product under several legislative works. Imbued with the results of the evaluation stage, the ethics board frames the product into the pre-established impact levels, giving rise to a particular set of recommendations (WHY–SHOULD–HOW). The EaaS platform delivers this documentation back to the organization, together with their tailored review. This deliverable presents criteria and tools to improve a product according to the principles under consideration. To further help, these tools are offered in a pedagogical form, i.e., via documented examples of use cases (e.g., how to perform adversarial exploration, evaluate fairness metrics, interpret language models, etc.), to improve their adoption and use. These are presented as step-by-step procedures to improve the organization's product further.

The workflow of our implementation combines aspects related to ethical principles, legal compliance, and technical implementations, articulating all of them together akin to the "Stronger Together" proposal of Pistilli et al. [117]. This, we argue, ultimately leads to tightening the principle–practice gap, i.e., from AI to beneficial AI. Meanwhile, the point in which our framework goes beyond past works in shortening the principle–practice gap lies heavily in our pedagogical aspect. Past frameworks almost always give you the normative (the ought) and, more rarely, the practical (the how). Besides giving normative recommendations, our framework seeks to teach developers how to use tools to tackle ethical and safety problems. At the same time, the materials that support our framework are all openly available, making this study one of the first efforts to tailor an AI governance approach to the Brazilian context in an open-source fashion while also allowing for spin-offs tailored to different contexts. Readers can access the source code and all other materials tied to the EPS framework at the following URL https://nkluge-correa.github.io/ethical-problem-solving/.

## 5 Limitations and future works

EPS represents a novel attempt to bridge the principle–practice gap between ethical principles and the practical implementation of AI development. However, much work remains.

First, while the framework provides a structured approach to addressing ethical concerns, handling AI's ever-evolving landscape is a tiring feature we must come to terms with. Many foundations of our approach rest on work that is bound to be outdated. The axiological roots of the EPS framework (the WAIE review), the legislative sources we used to create our impact assessment surveys, our risk assessment catalog, and the practices we recommend and exemplify are all bound to become irrelevant as advances in these fronts occur. Hence, the fast-moving nature of the field requires implementations of our framework to undergo constant recycling; otherwise, we risk falling into uselessness. This fact leads us to the question, "Can humane sciences accompany the accelerated pace of the technological industry?" which, in our opinion, has till now been answered negatively and with pessimism. As mentioned, bridging the principle–practice gap is a continuous problem-solving process, and as Jaeggi points out [52], problem-solving is a never-ending work. Hence, an obvious future avenue of work involves updating and extending EPS. For example, there is undoubtedly space to expand the legislative contribution in subsequent implementations of the EPS, even more so if this expansion encompasses legislation specifically focused on AI systems (e.g., Generative AI). Also, it remains open the possibility to integrate more general frameworks, like international human rights [118–120], which are already a part of some impact assessment tools tailored to the assessment of human rights [121].

Another sensitive point of our framework regarding its evaluation method is its scoring process. In our current implementation, we constrained our impact score scale to a standard range, where answers to questions could maximally impact a given principle with a score of 1 or decrease its impact with a score of − 1. At the same time, we chose to have more ways in which impact scores could be increased rather than decreased, i.e., $\approx \frac{1}{5}$ of the general score from each AIA produces a reduction in impact, given that the main objective of our evaluations is to assess the lack of ethical compliance. Hence, one issue we face is the feasibility of translating regulatory standards into a cardinal evaluation scale. The problem of intertheoretic comparisons and the gauging of how much "utility" a developmental choice should represent is not trivial, being an area of open research in Moral Philosophy and Metanormativity [122–125]. Given that we developed these evaluations to inform a body of experts imbued with making a normative framing, finding ways to present evaluations understandably and unambiguously is crucial. While our approach translates standardized cardinal values to unambiguous impact classes (low, intermediate, and high), other methods might be better suited. Searching for improved ways to perform this translation is an area of study worth pursuing.

Meanwhile, the idea of a human-centric evaluation presents its own problem. This human-centric aspect, which, in the end, comes down to the biased and subjective view of a group of individuals, is one of its weak spots. Like many other forms of evaluation and certification that rely on human oversight, the EPS may also fall short of its promise if its human element is unaligned with the core purpose of the framework. While the idea of an EaaS platform that should always be managed (or at least audited) by an external party, such as a government body or a neutral auditing organization, may help avoid specific perils without a proper normative engine (a.k.a. an evaluation board or an ethics committee), the whole idea of EaaS could deteriorate into *Ethics Washing as a Service* [126]. We remain committed to the concept of not automatizing ethics. However, we argue that the success of this type of framework also rests in the question of how to train and educate good ethical boards to perform this crucial role [109, 110], which is another avenue for future studies.

This work also explored the limitations of shortening the principle–practice gap with a toolbox mentality. In other words, we encountered several cases where we may need more than mere tools and the knowledge of how to use them to fulfill our higher goals. For example, one can use statistical metrics and other tools to assess the fairness of an AI system and further correct them. However, these do not attack the root cause of inequality that becomes imprinted in our systems and applications [127]. One can use several methods to protect and obfuscate personally identifiable information during an AI development cycle. Yet, robust privacy can only be achieved through collaboration among governments, private companies, and other stakeholders [128, 129]. One can use carbon tracking to offset their footprint and promote sustainable practices. But unfortunately, sustainability in AI ethics cannot be reduced to such mere calculations, given the many other side effects of our technological progress do not have an easy-to-measure metric, e.g., the depletion of our natural resources [130], the humanitarian costs related to their extraction [5, 131, 132], production of e-waste [133], etc. All these cases stress that AI ethics has challenges beyond the "lack of implementational techniques" or "knowledge gaps," which should incentivize works and agendas that use a different approach than the one we utilized in the EPS.

Other limitations that fall outside the scope of our framework but can prevent (or improve) its success are these:

- **Regulatory supports and incentives:** frameworks like the EPS and the development of EaaS platforms can become a future necessity if regulatory bodies make this evaluation a mandatory procedure for AI products above a certain impact level. At the same time, regulatory bodies could adopt frameworks like this and serve it as a certification system. On the other hand, it could also be the case that frameworks like these would only have adoption with regulatory support and, without it, would find no adoption in the industry. Just as in the case of organizations that provide cybersecurity and GDPR compliance services, their adoption is tied to regulations that make compliance necessary for them.
- **Lack of attention to ethical issues in entrepreneurial environments:** as already mentioned by previous studies [114], entrepreneurial settings, where many modern technologies become commoditized into products, are not environments that usually take ethical concerns too seriously, where these are generally seen as a nuisance or barrier to further progress. Currently, the field necessitates minimizing knowledge asymmetry between all sectors, from humane sciences to STEM fields. To research and develop business and applications. However, there are still many obstacles to this type of collaboration and how to overcome the challenges of interdisciplinary research.
- **Lack of virtues in AI and software developers:** studies have already shown that we might have a Humane Sciences gap in STEM areas [113, 134], while questions related to AI ethics and safety are still far from the mainstream in terms of Academic curricula. However, we argue that practices like the ones promoted in our framework could help patch this in the educational development of STEM professionals acting as AI engineers and developers. Regardless, improving the "Humane aspect" of the formation of these professionals could undoubtedly improve their sensibility to the issues dealt with by frameworks like the EPS.

These are research areas that can, directly and indirectly, improve not just the success of this study's objectives, i.e., shortening the principle–practice gap, but AI Ethics itself. In this landscape, our framework is a blueprint for other researchers to build upon and expand. The entirety of the proposed process gives more than enough space for the many areas related to AI ethics to contribute, like, for example, improving evaluation methods, coming up with recommendations for ethical design choices, creating tools, or teaching developers how to use them. Our objective for the foreseeable future is to fully implement and test the EPS framework as an EaaS platform in Brazil while supporting and updating our open-source repositories. We hope this work and service may provide novel pathways and opportunities to AI ethicists and better general guidance and assistance to the field.

## 6 Conclusion

In this work, we presented ethical problem-solving, a broad framework pursuing the betterment of AI systems. Our framework employs impact assessment tools and a differential recommendation methodology. The EPS framework differentiates itself from other works by its theoretical grounding, axiological foundation, and operationalization, serving as the blueprints for an EaaS platform that mediates the normative expectations of the field and the reality of AI system development. However, crossing the principle–practice gap in AI development is an ongoing process. Even though many problems remain without immediate technical solutions, efforts like this can help institute a culture of responsible AI practice in a way that can keep pace with the advances in the field. Finally, by opening our work and results, we hope other researchers can easily extend and surpass our work.

**Acknowledgements** This research was funded by RAIES (Rede de Inteligência Artificial Ética e Segura). RAIES is a project supported by FAPERGS (Fundação de Amparo à Pesquisa do Estado do Rio Grande do Sul) and CNPq (Conselho Nacional de Desenvolvimento Científico e Tecnológico).

**Author Contributions** This study involved a collaborative effort from a multidisciplinary team. N.K.C. contributed to the development of the methodology, the creation of the demo and related code repositories, and the writing of the article. J.W.S. contributed to the development of the methodology, the documentation of the demo, and the writing of the article. C.G. contributed to the development of the AIAs, to writing the corresponding sections related to the AIAs, and aided in the development of the overall methodology. M.P. contributed to the development of the AIAs, to writing the corresponding sections related to the AIAs, and aided in the development of the overall methodology. D.S. contributed to the development of code repositories, to the writing of the article, and aided in the development of the overall methodology. F.N. contributed to the development of the recommendation approach and aided in the development of the overall methodology. R.H. contributed to the development of the recommendation approach and aided in the development of the overall methodology. N.O. is the project coordinator.

**Funding** Open Access funding enabled and organized by Projekt DEAL.

**Data availability** The authors confirm that all data related to this study is available at the following URL: https://github.com/Nkluge-correa/ethical-problem-solving.

**Declarations**

**Conflict of interest** The authors declare no conflict of interest.